# Simulation of the 2E Technique on neutron multiplicity measurement as a function of fragment mass in spontaneous fission of $^{252}$Cf

M. Montoya

Universidad Nacional de Ingeniería, Av. Túpac Amaru 210, Rímac, Lima, Peru.

**Abstract**

A Monte Carlo simulation algorithm to investigate the measurement of the average prompt neutron multiplicity as a function of pre-neutron mass, $\bar{\nu}(A)$, for fragments from the spontaneous fission of $^{252}$Cf is presented. The input data consist of experimental measurements of the kinetic energy and mass distributions obtained by Göök et al. and the values of $\bar{\nu}(A)$ calculated using the FIFRELIN model by Piau et al., $\bar{\nu}_{\text{th}}(A)$. We analyze the output curve, $\bar{\nu}_{\text{sim}}(A)$, obtained by simulation of the 2E technique, which should ideally match $\bar{\nu}_{\text{th}}(A)$. However, we find that $\bar{\nu}_{\text{sim}}(A)$ exhibits a maximum value at $A \approx 122$, close to mass symmetry, while $\bar{\nu}_{\text{th}}(A)$ has a maximum at $A \approx 118$. Additionally, we observe that que $\bar{\nu}_{\text{sim}}(A) > \bar{\nu}_{\text{th}}(A)$ for $A < 90$ and $A > 162$, respectively. We attribute this discrepancy to inaccuracies in the relationship between provisional mass and the pre-neutron mass used in the 2E technique for data processing in each fission event.

## 1 Introduction

The fission process begins with the fissile nucleus, composed of $A_0$ nucleons, where $Z_0$ are protons, and the remaining are neutrons. Nucleons interact through nuclear and Coulomb interactions. The fission process culminates at the scission point, where the nuclear interaction fades, and two complementary fragments with proton and neutron numbers $(Z, A)$ and $(Z', A')$, and excitation energies $(XE, XE')$, respectively, are formed [1] [2], [3]. Studying the dynamics of fission requires knowledge of the distribution of these variables, which is not directly accessible due to the dimensions at which the phenomenon occurs.

An intermediate step would involve measuring the distribution of kinetic energy values acquired by the fragments due to the Coulomb repulsion they experience after the scission point $(E, E')$. Additionally, it is essential to measure the excitation energy of the fragments, $XE$ and $XE'$. With these values, the available energy of the reaction can be calculated:

$$Q = TKE + TXE, \qquad (1)$$

where $TKE = E + E'$.

However, for each fission event, the fragments arrive at the detectors after having emitted immediate neutrons in numbers $(n, n')$, respectively, of which only the average has been calculated. Thus, the post-neutron values of the kinetic energy of the fragments $(e, e')$, respectively, result from mass loss and recoil effects due to neutron emission. The recoil effect from the emission of all neutrons is not fully accessible, lacking access to all necessary variables. For each neutron emission, it would be necessary to measure its kinetic energy and emission direction.

With these limitations, Göök et al. have employed the 2E technique to measure the distribution of $A$, $TKE$ $(= E + E')$, and the average prompt neutron multiplicity as a function of fragment mass, denoted as $\bar{\nu}_{\text{ex}}(A)$ [4]. On the other hand, there are results from calculations based on theoretical models that compute the values of $\bar{\nu}(A)$, referred to as $\bar{\nu}_{\text{th}}(A)$. Among these is the FIFRELIN model used by Piau et al. [5].

In this study, it is employed a simulation algorithm for fission experiments to investigate the disparity between the results obtained through the $2E$ measurement technique and the values of the average prompt neutron multiplicity as a function of pre-neutron mass, $\bar{\nu}(A)$, assumed as real, for fragments from the spontaneous fission of $^{252}$Cf.

## 2 Monte Carlo simulation of de $\bar{\nu}(A)$ measurement using the $2E$ technique

As input data corresponding to the primary fragments with mass $A$ and energy $E$, the values measured by Göök et al. is considered[4]: *i)* the yield $Y(A)$; *ii)* the average kinetic energy $\overline{TKE}(A)$ and its standard deviation $\sigma_{TKE}(A)$; *iii)* the partial derivative of $\bar{\nu}$ with respect to $TKE$, $\partial \bar{\nu}/\partial TKE$; and *iv)* the average kinetic energy of the emitted immediate neutrons $\bar{\eta}(A)$. The values of the average neutron multiplicity we use as input data are those calculated by Piau et al. [5] $\bar{\nu}_{\text{th}}(A)$.

For the simulation, we assume a Gaussian distribution of $TKE(A)$:

$$TKE = \overline{TKE} + r\sigma_{TKE}(A), \qquad (2)$$

where $r$ is a number that follows a Gaussian distribution with a mean of 0 and a standard deviation of 1.

For the neutron multiplicity, we employ the following approximate relationship:

$$\bar{\nu}_{\text{th}}(A, TKE) = \bar{\nu}_{\text{th}}(A)\left(1 + \frac{\partial \bar{\nu}}{\partial TKE}(\overline{TKE} - TKE)\right). \qquad (2)$$

The simulation involves the emission of an integer number $n$ of neutrons, with an average $\bar{\nu}_{\text{th}}(A, TKE)$. As the standard deviation data is unavailable, we assume that fragments with mass $A$ and total kinetic energy $TKE$ emit the integer number closest to $\bar{\nu}_{\text{th}}$.

Additionally, it is assumed that fragments with mass $A$ associated with the total kinetic energy $TKE$ emit isotropically n neutrons with kinetic energy $\bar{\eta}(A)$. The recoil effect due to neutron emission is then calculated to obtain the final kinetic energy values of the fragments $e_i$ ($i = 1,2$). Based on these values, the provisional mass of the two fragments is calculated:

$$\mu_{1,2} = 252\left(\frac{e_{2,1}}{tke}\right), \qquad (3)$$

where $tke = e_1 + e_2$.

## 3. Results

For the input values of the Monte Carlo simulation, we consider the $\bar{\nu}_{\text{th}}(A)$ values calculated by Piau et al. [5].. The outcome of the Monte Carlo simulation for the $2E$ technique measurement of $\bar{\nu}$ as a function of provisional mass, $\bar{\nu}_{\text{prov}}(\mu)$, is presented in Figure 1.

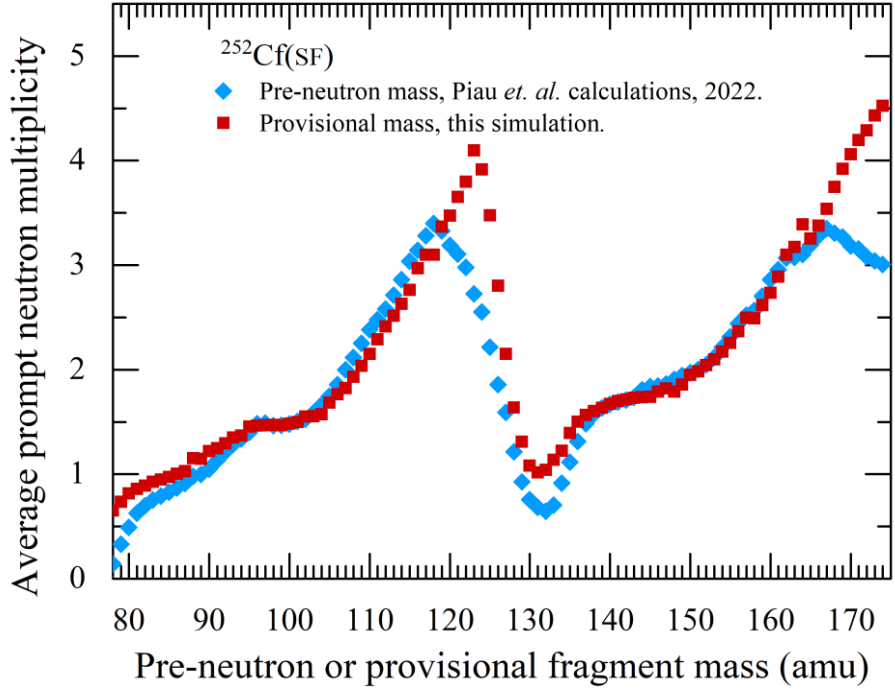

Figure 1: Average prompt neutron multiplicity as a function of pre-neutron mass, calculated by Piau et al. ($\bar{\nu}_{\text{th}}$: diamonds) [5], and as a function of provisional mass ($\bar{\nu}_{\text{prov}}$: squares) for fragments from the spontaneous fission of $^{252}$Cf, obtained as the output of the Monte Carlo simulation.

To approximately calculate the pre-neutron values of kinetic energy, we utilize the following relationship:

$$E_i \approx tke\left(\frac{\mu_i}{\mu_i - \bar{\nu}_{\text{prov}}(\mu_i, tke)}\right). \tag{4}$$

This relation is an approximation of the formula:

$$e_i = TKE\left(\frac{A_i}{A_i - n_i}\right). \tag{5}$$

which relates the pre-neutron kinetic energy to the final energy of fragments emitting $n_i$ neutrons with zero kinetic energy in the center of mass frame.

From the approximate values of kinetic energy obtained using equation (4), we calculate the approximate values of the primary fragment masses:

$$A_{1,2} \approx 252\left(\frac{E_{2,1}}{TKE}\right), \tag{6}$$

With the values of $n$ corresponding to each mass $A$, the $\bar{\nu}_{\text{sim}}(A)$ curve is constructed and presented in Figure 2.

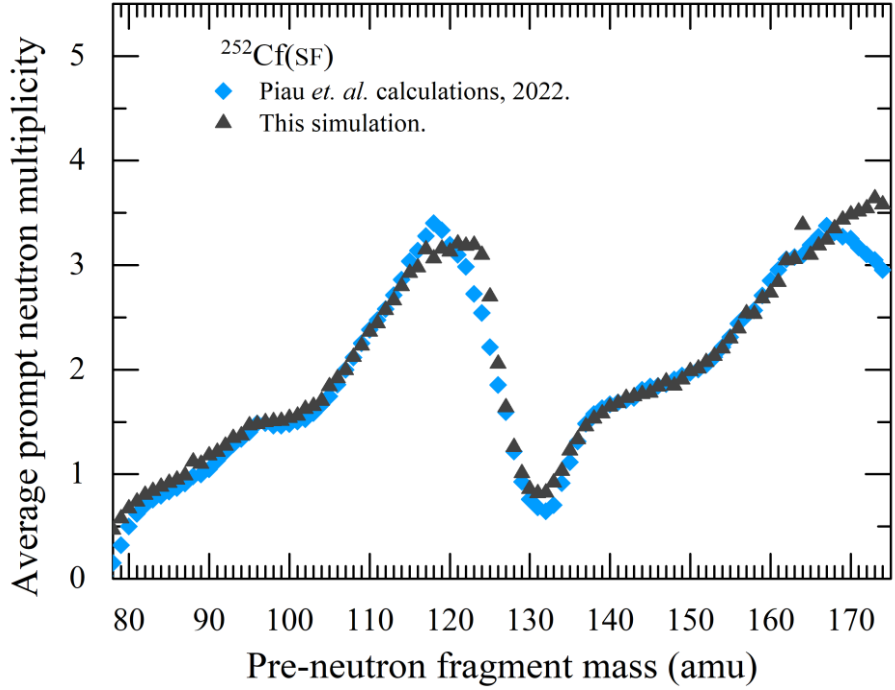

Figure 2: Average prompt neutron multiplicity as a function of pre-neutron mass, calculated by Piau et al. ($\bar{\nu}_{\text{th}}$: diamonds) [5], and as a function of pre-neutron mass for fragments from spontaneous fission of $^{252}$Cf, simulated using the $2E$ technique algorithm ($\bar{\nu}_{\text{sim}}(A)$: triangles).

Fig. 3 presents the curves $\bar{\nu}_{\text{th}}(A)$, $\bar{\nu}_{\text{sim}}(A)$, and the experimental curve $\bar{\nu}_{\text{ex}}(A)$, obtained by Göök et al. [4]. The curves $\bar{\nu}_{\text{sim}}(A)$ and $\bar{\nu}_{\text{ex}}(A)$ exhibit similar trends, except for the region $A < 80$, where $\bar{\nu}_{\text{ex}}(A)$ greatly surpasses $\bar{\nu}_{\text{sim}}(A)$. Both curves reach a maximum value at $A = 122$, which is 4 units higher than the corresponding maximum of $\bar{\nu}_{\text{th}}(A)$. Additionally, both curves have higher values than $\bar{\nu}_{\text{th}}(A)$ for $A > 169$.

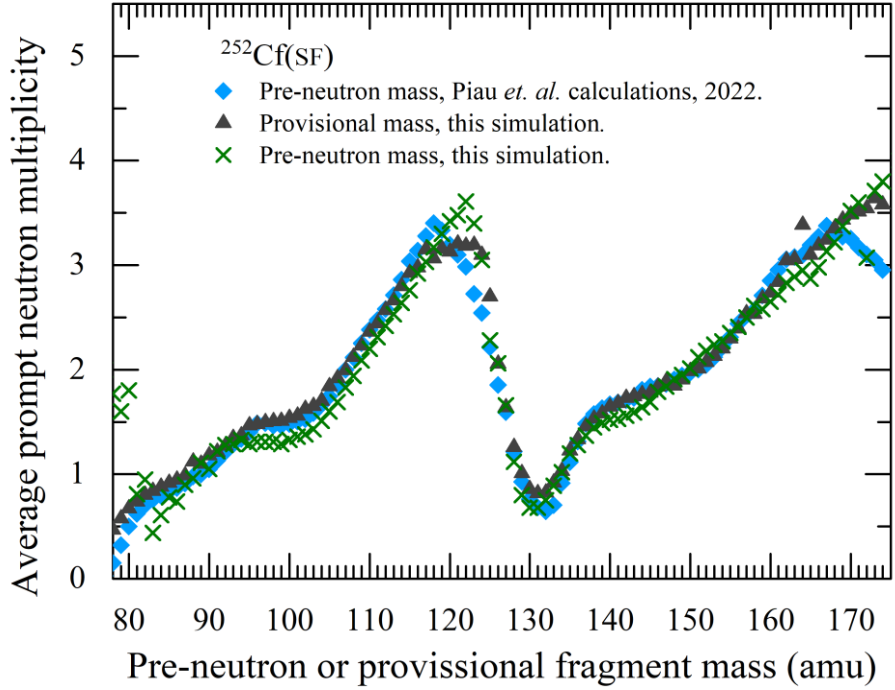

Figure 3: Average prompt neutron multiplicity as a function of pre-neutron mass, calculated by Piau et al. ($\bar{v}_{th}$: diamonds) [5], for fragments from spontaneous fission of $^{252}$Cf, simulated using the $2E$ technique algorithm ($\bar{v}_{sim}(A)$: triangles), and the experimental curve measured by Göök et al. ($\bar{v}_{exp}(A)$ : blades) [4].

## 4. Conclusion

In this study, the average prompt neutron multiplicity as a function of provisional and pre-neutron masses of fragments from the spontaneous fission of $^{252}$Cf was simulated using the Monte Carlo method with the $2E$ technique. Fig. 1 illustrates that the variable $\bar{v}_{prov}(\mu)$, concerning provisional mass, exhibits a peak at $\mu = 122$, whereas the theoretical primary mass-related variable, $\bar{v}_{th}(A)$, reaches a maximum at $A = 118$. Moreover, $\bar{v}_{th} < \bar{v}_{prov}$ for A<90 and A>162. The simulation output of $\bar{v}_{sim}(A)$ with respect to primary mass ($A$) reflects the same characteristics as $\bar{v}_{th}$.

The discrepancy between the simulated curve $\bar{v}_{sim}(A)$, generated through the $2E$ technique simulation, and the input curve $\bar{v}_{th}$ is attributed to the inexactness of the relationship (4), which connects provisional mass and pre-neutron energy. A better approximation could be achieved by considering the number of neutrons ($n$) emitted by the fragment instead of $\bar{v}_{prov}(\mu_i, tke)$, representing an average neutron distribution. However, even with this refinement, the relationship would still be imprecise, as it does not account for the recoil effect caused by neutron emission, which alters the kinetic energy of the emitting fragments.